%% file: paper_hepex.tex
\documentclass[twocolumn,showpacs,aps,prl,superscriptaddress,floatfix]{revtex4}
\usepackage{graphicx}
\usepackage{dcolumn}
\usepackage{multirow}
\usepackage{amsmath}
\usepackage{epsfig}
\usepackage{color}

\input pubboard/babarsym

\newcommand{\BABARPubYear}    {05}
\newcommand{\BABARPubNumber} {025}
\newcommand{\SLACPubNumber}{11053}

\def\rhopm{\ensuremath{\rho^{\pm}}\xspace}
\def\rhoz{\ensuremath{\rho^{0}}\xspace}
\def\hu{\ensuremath{h_u}\xspace}
\def\Xu{\ensuremath{X_u}\xspace}
\def\Bpirholnu{\ensuremath{B \rightarrow \pi(\rho) \ell\nu}\xspace}
\def\Bzpilnu{\ensuremath{B^{0} \rightarrow \pi^-\ell^+\nu}\xspace}
\def\Bppizlnu{\ensuremath{B^{+} \rightarrow \pi^0\ell^+\nu}\xspace}
\def\Bzrholnu{\ensuremath{B^{0} \rightarrow \rho^-\ell^+\nu}\xspace}
\def\Bprhozlnu{\ensuremath{B^{+} \rightarrow \rho^0\ell^+\nu}\xspace}
\def\Bpomegalnu{\ensuremath{B^{+} \rightarrow \omega \ell^+\nu}\xspace}
\def\Bpetalnu{\ensuremath{B^{+} \rightarrow \eta \ell^+\nu}\xspace}

\def\pilnu{\ensuremath{\pi\ell\nu}\xspace}
\def\pimlnu{\ensuremath{\pi^-\ell^+\nu}\xspace}
\def\pizlnu{\ensuremath{\pi^0\ell^+\nu}\xspace}
\def\rholnu{\ensuremath{\rho\ell\nu}\xspace}
\def\rhomlnu{\ensuremath{\rho^-\ell^+\nu}\xspace}
\def\rhozlnu{\ensuremath{\rho^0\ell^+\nu}\xspace}
\def\BHulnu{\ensuremath{B \rightarrow h_u\ell\nu}\xspace}
\def\BXulnu{\ensuremath{B \rightarrow X_u\ell\nu}\xspace}

\def\bulnu{\ensuremath{B \rightarrow X_u\ell\nu}\xspace}
\def\bclnu{\ensuremath{B \rightarrow X_c\ell\nu}\xspace}

\def\BRBzpilnu{\ensuremath{{\cal B}(B^{0} \rightarrow \pi^-\ell^+\nu})\xspace}
\def\BRBppizlnu{\ensuremath{{\cal B}(B^{+} \rightarrow \pi^0\ell^+\nu})\xspace}
\def\BRBzrholnu{\ensuremath{{\cal B}(B^{0} \rightarrow \rho^-\ell^+\nu})\xspace}

\def\mES{\ensuremath{m_{\rm ES}}\xspace}
\def\DeltaE{\ensuremath{\Delta E}\xspace}
\def\cosBYDef{\ensuremath{\cos \theta_{BY} = \left(2 E^*_B
      E^*_Y-M_B^2-M_Y^2\right)/\left(2|\vec p^{\,*}_B|
      |\vec p^{\,*}_Y|\right)}\xspace}
\def\mESDef{\ensuremath{m_{\rm ES} = \sqrt{(s/2+\vec{p}_B \cdot \vec{p}_{\rm beams})^2/E_{\rm beams}^2- \vec{p}_B^{\,2}}}\xspace}
\def\DeltaEDef{\ensuremath{\Delta E = (p_B \cdot p_{\rm beams} - s/2) / \sqrt{s}}\xspace}
\def\qsquaredDef{\ensuremath{q^2 = (p_\ell + p_\nu)^2}}

\def\ThetamissCut{\ensuremath{0.6 < \theta_{\rm miss} < 2.9} \rm rad\xspace}

\def\plepCut{\ensuremath{|\vec{p}^{\,*}_{\rm \ell}| > 1.3~\gev}\xspace}
\def\pstarlpstarhPiCut{\ensuremath{|\vec{p}^{\,*}_\ell|+|\vec{p}^{\,*}_{h_u}| > 2.6 \gev}\xspace}
\def\pstarlpstarhRhoCut{\ensuremath{1.5 |\vec{p}^{\,*}_\ell|+|\vec{p}^{\,*}_{h_u}| > 4.2 \gev}\xspace}
\def\pstarlRhoCut{\ensuremath{|\vec{p}^{\,*}_\ell| > 1.8 \gev }\xspace}
\def\mmissEmissCut{\ensuremath{|m_{\rm miss}^2 / 2 E_{\rm miss} | < 0.4~\gev }\xspace}
\def\cosBYCut{\ensuremath{|\cos \theta_{\rm BY}| < 1.1}\xspace}

\def\FitRegion{\ensuremath{|\Delta E| < 0.9~\gev \mbox{ and } m_{\rm ES} > 5.095~\gev}\xspace}

\def\SignalBandDeltaE{\ensuremath{-0.15 < \Delta E < 0.25~\gev}\xspace}
\def\SignalBandmES{\ensuremath{m_{\rm ES} > 5.255~\gev}\xspace}
\def\fpmfzz{\ensuremath{f_{+-}/f_{00} = 1.044 \pm 0.050}\xspace}
\def\BlifetimeRatio{\ensuremath{\tau_{B^\pm}/\tau_{B^0} = 1.081 \pm 0.015}\xspace}

\begin{document}

\preprint{\babar-PUB-\BABARPubYear/\BABARPubNumber} 
\preprint{SLAC-PUB-\SLACPubNumber} 

\begin{flushleft}
\babar-PUB-\BABARPubYear/\BABARPubNumber\\
SLAC-PUB-\SLACPubNumber\\
Published in Phys. Rev. D {\bf 72}, 051102 (2005)\\
\end{flushleft}

\begin{flushright}
\end{flushright}

\vspace*{0.6cm}

\title{\large\bf Study of \boldmath \Btopilnu and \Btorholnu decays and determination of \Vub}

\input pubboard/authors_pub05025_weinstein.tex

\date{\today}

\begin{abstract}
\input abstract.tex

\end{abstract}

\pacs{13.20.He,                 
      12.15.Hh,                 
      12.38.Qk,                 
      14.40.Nd}                 

\maketitle  


The parameter \Vub is one of the smallest and least known
elements of the Cabibbo-Kobayashi-Maskawa (CKM) quark-mixing matrix~\cite{ckm}.
A precise determination of \Vub would significantly improve the constraints 
on the Unitarity Triangle and provide a stringent test of the Standard
Model mechanism for \CP violation.
In this paper, we present a determination of \Vub from charmless
semileptonic decays of \B mesons with exclusively reconstructed final states,
\BHulnu,
where the hadronic state \hu represents a \pipm, \piz, \rhopm, or \rhoz and
$\ell$ represents $e$ or $\mu$. 
Exclusive decays allow for kinematic constraints and more efficient 
background suppression compared to inclusive decays, but must rely on theoretical
form-factor predictions. 
Using isospin symmetry, we measure the branching fractions 
\BRBzpilnu~\footnote{Charge-conjugate modes are included implicitly.}
and \BRBzrholnu\ as a function of \qsquaredDef, the momentum-transfer squared, 
and extract \Vub using recent form-factor calculations 
based on light-cone sum rules (LCSR)~\cite{lcsr:pi,lcsr:rho} and unquenched 
lattice QCD (LQCD)~\cite{lqcd:hpqcd,lqcd:fnal}. 

This measurement is based on a sample of 
83 million \BB\ pairs recorded with the \babar\ 
detector~\cite{babar} at the PEP-II asymmetric-energy \epem storage rings. 
The data correspond to an integrated luminosity of 75.6~\invfb
collected at the \FourS resonance and 8.9~\invfb recorded 40 \mev\ below it.
Simulated \BB\ events
are used to estimate signal efficiencies and shapes of signal and background
distributions. Charmless semileptonic decays are simulated as a mixture of 
three-body decays \BXulnu ($\Xu = \pi, \eta, \eta', \rho, \omega$) based on the 
ISGW~II quark model~\cite{isgw2}. 
Decays to non-resonant hadronic states \Xu with masses $m_{\Xu} > 2m_\pi$ 
are simulated following a prescription of Ref.~\cite{DFN}.

We identify charmless semileptonic decays by 
a charged lepton with momentum \plepCut\footnote{All variables
denoted with a star (e.g. $p^*$) are given in the \FourS rest frame; all
others are given in the laboratory frame.}, a $\pi$ or $\rho$ meson,
and missing momentum $|\vec p_{\rm miss}|>0.7~\gev$ in the event.
We identify $\rho$ mesons via the decays 
$\rho^\pm \rightarrow \pi^\pm \pi^0$ and $\rho^0 \rightarrow \pi^+\pi^-$
with mass  $0.65 < m_{\pi\pi}< 0.85\gev$, rejecting candidates in which a
charged track is identified as a kaon; both \pipm and $\rho$ candidates
are rejected if a charged track is identified as a lepton.
The charged lepton is combined with a \piz, \rhoz or \pipm, \rhopm 
of opposite charge to form a ``$Y$'' candidate; $Y$ candidates
are rejected if the lepton and an oppositely-charged track
from the signal hadron are consistent with a $J/\psi \rightarrow \ell^+\ell^-$ decay.

The neutrino four-momentum, $p_{\nu}=(E_{\rm miss},\vec{p}_{\rm miss})$, is inferred from the
difference between the net four-momentum of the colliding-beam particles,
$p_{\rm beams}=(E_{\rm beams},\vec{p}_{\rm beams})$,
and the sum of the four-momenta of all detected particles in the event.
To reduce the effect of losses due to the detector
acceptance, we require a total charge of the event of $|Q_{\rm tot}|\leq 1$ 
and a polar angle of the missing momentum in the range \ThetamissCut.
In addition, the missing mass measured from the whole event should be compatible 
with zero.  Because the missing-mass resolution varies
linearly with the missing energy, we require \mmissEmissCut .
We compute the angle between the $Y$ candidate and the \B\ meson,
assuming zero missing mass, as \cosBYDef.  
Here $M_B, M_Y, E^*_B, E^*_Y, \vec p^{\,*}_B, \vec p^{\,*}_Y$ refer to the 
masses, energies, and momenta of the \B\ and $Y$.
Signal candidates are required to satisfy \cosBYCut, 
allowing for detector resolution and photon radiation.

We restrict the momenta of leptons and hadrons in $Y$ candidates to enhance the 
signal over backgrounds.  For \Btopilnu, we require \pstarlpstarhPiCut; 
for \Btorholnu, \pstarlpstarhRhoCut and \pstarlRhoCut. 
These criteria keep 99.8\% (75\%) of true \Bpirholnu
decays and reduce the \bclnu background by about 10\% (80\%) after all other
selection criteria.  To suppress backgrounds from $e^+e^- \ra \qqbar\,
(q=u,d,s,c)$ and QED processes,
we require at least five charged tracks in each event or, to increase
the efficiency for \Bppizlnu, four tracks and at least two photons.  
We also require $L_2 = \sum_i |\vec{p}^{\,*}_i|
{\cos}^2\theta^*_i < 1.5$~\gev. Here the sum is over 
all tracks in the event excluding the $Y$ candidate, and $\vec{p}^{\,*}_i$ and $\theta^*_i$ 
refer to the momenta and the angles measured with respect to the thrust axis 
of the $Y$.  This requirement removes over 95\% of \qqbar\ and 80\% of \bclnu\ 
background and retains about 50\% of the signal in all modes.

We discriminate against the remaining back\-ground using the variables
\DeltaEDef and \mESDef,
where $\sqrt{s}$ is the mass of the \FourS.
Only candidates with \FitRegion\ are retained. The total
signal selection efficiencies for the sum of electrons and muons are
3.5\% and 2.4\% for \pimlnu\ and \pizlnu, 0.53\% and 1.1\% 
for \rhomlnu\ and \rhozlnu.~\footnote{These efficiencies
and signal yields depend upon $q^2$-dependent form factors which, unless
otherwise stated, are fit to
the data for \Btopilnu and calculated using LCSR~\cite{lcsr:rho} for
\Btorholnu.}
We use a low-background sample of $B^0 \ra D^{*-} \ell^+ \nu$ decays with 
$\Dbar^0 \ra K^+ \pi^-$ or $\Dbar^0 \ra K^+ \pi^- \pi^0$ to compare the efficiencies 
of each selection cut in data and simulation and find differences typically of a few percent.

To extract the signal yields, we perform a binned extended
maximum-likelihood fit~\cite{bbfit} to the \DeltaE vs. \mES distributions 
of the four signal modes simultaneously.  
The fit takes into account statistical
fluctuations of both data and Monte-Carlo samples.  We fit the
relative proportions of the simulated signal and background samples to the
data distributions in $5 \gev^2$ or $10 \gev^2$ intervals of $q^2$. 
To improve the $q^2$ resolution, 
we adjust $|\vec p_{\nu}|$ so that $\DeltaE =0$. 
The resulting $q^2$ resolution is small compared to the chosen intervals of $q^2$
and can be described by
the sum of two Gaussian functions of widths $\sigma_1 \simeq 0.2\gev^2$ 
(containing about 75\% of signal events) and $\sigma_2 \simeq 0.5\gev^2$.

We use the isospin relations
$\Gamma(\Bzpilnu) = 2\Gamma(\Bppizlnu)$ and 
$\Gamma(\Bzrholnu) = 2\Gamma(\Bprhozlnu)$
to reduce the number of fit parameters to nine: five for the signal
yields in
the five $q^2$ intervals for $B \ra \pi \ell \nu$ decays, three for the
signal yields in the three $q^2$ intervals for $B \ra \rho \ell \nu$ decays, plus
one scale parameter, shared among all $q^2$ intervals and signal modes, to fit the
overall normalization of the \bclnu background. 
We classify signal candidates as ``combinatoric signal'' if the reconstructed lepton
comes from the isospin-conjugate decay or the hadron is incorrectly selected. 
The fit uses common parameters for combinatoric signal and signal. 
The normalization of the simulated non-\BB\ background is scaled 
separately for events with $e^{\pm}$ and $\mu^{\pm}$ 
to match the off-resonance data. We smooth the distributions
for this low-statistics background to reduce single-bin statistical fluctuations.

\input{tab_bfq2.tex}

\begin{figure*}[htbp]
\begin{center}
\epsfig{file=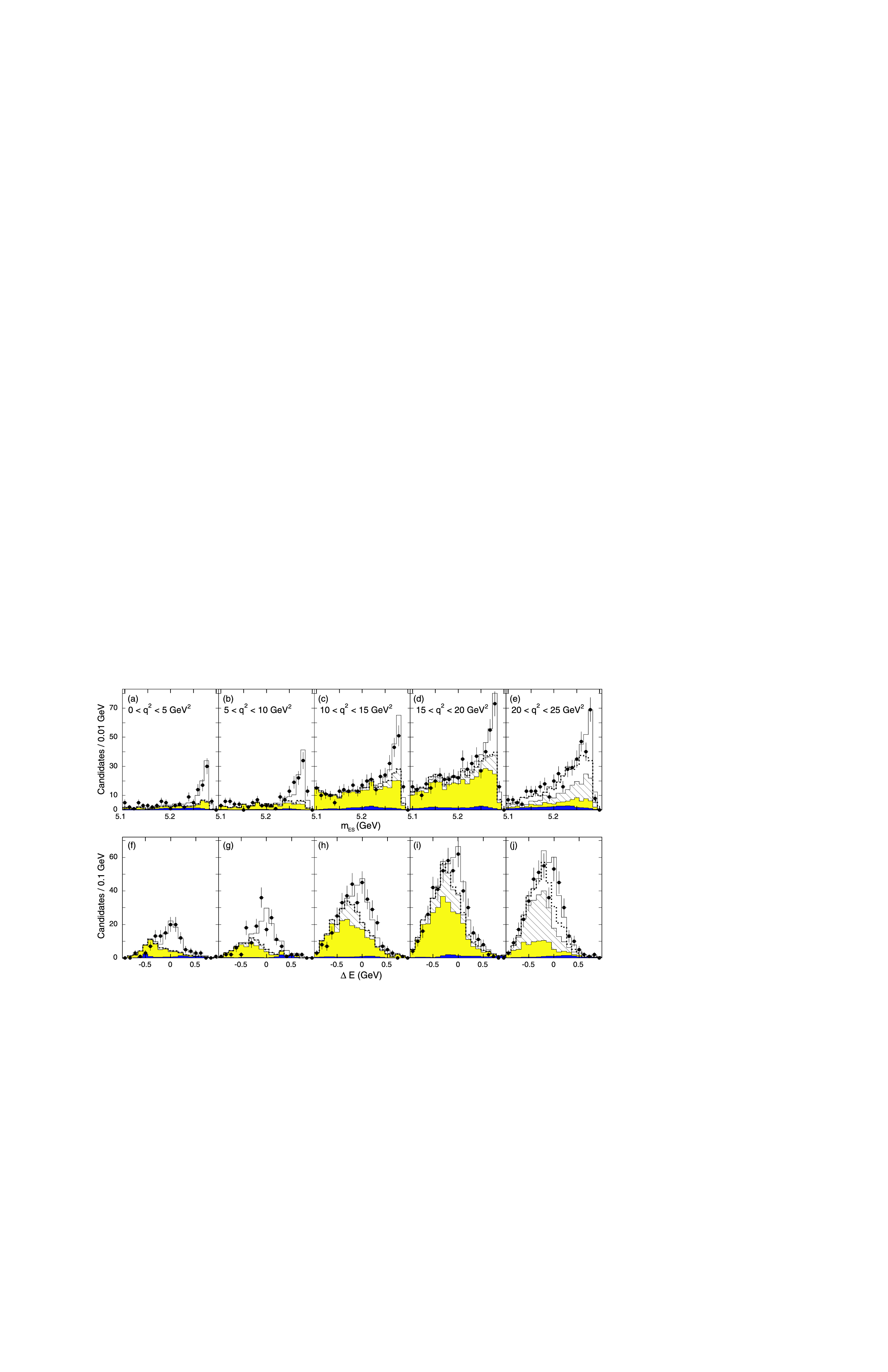, width = 15cm}
\caption{(color online) Projected \mES (a-e) and \DeltaE (f-j) distributions 
in five intervals of $q^2$ for the combined \Btopilnu modes. 
The projections are shown for signal bands \SignalBandDeltaE and \SignalBandmES, respectively.
The error bars on the data points represent the statistical uncertainties.
The histograms show simulated 
distributions for signal (white), combinatoric signal (white, dotted), 
cross feed from other \bulnu decays (hatched), \bclnu decays (light shaded/yellow) 
and non-\BB\ background (dark shaded/blue). The normalizations of the signal and \bclnu background 
simulations have been scaled to the results of the maximum-likelihood fit.
}
\label{fig:MesDeltaE:pi}
\end{center}
\end{figure*}

\begin{figure}[htbp]
\begin{center}
\epsfig{file=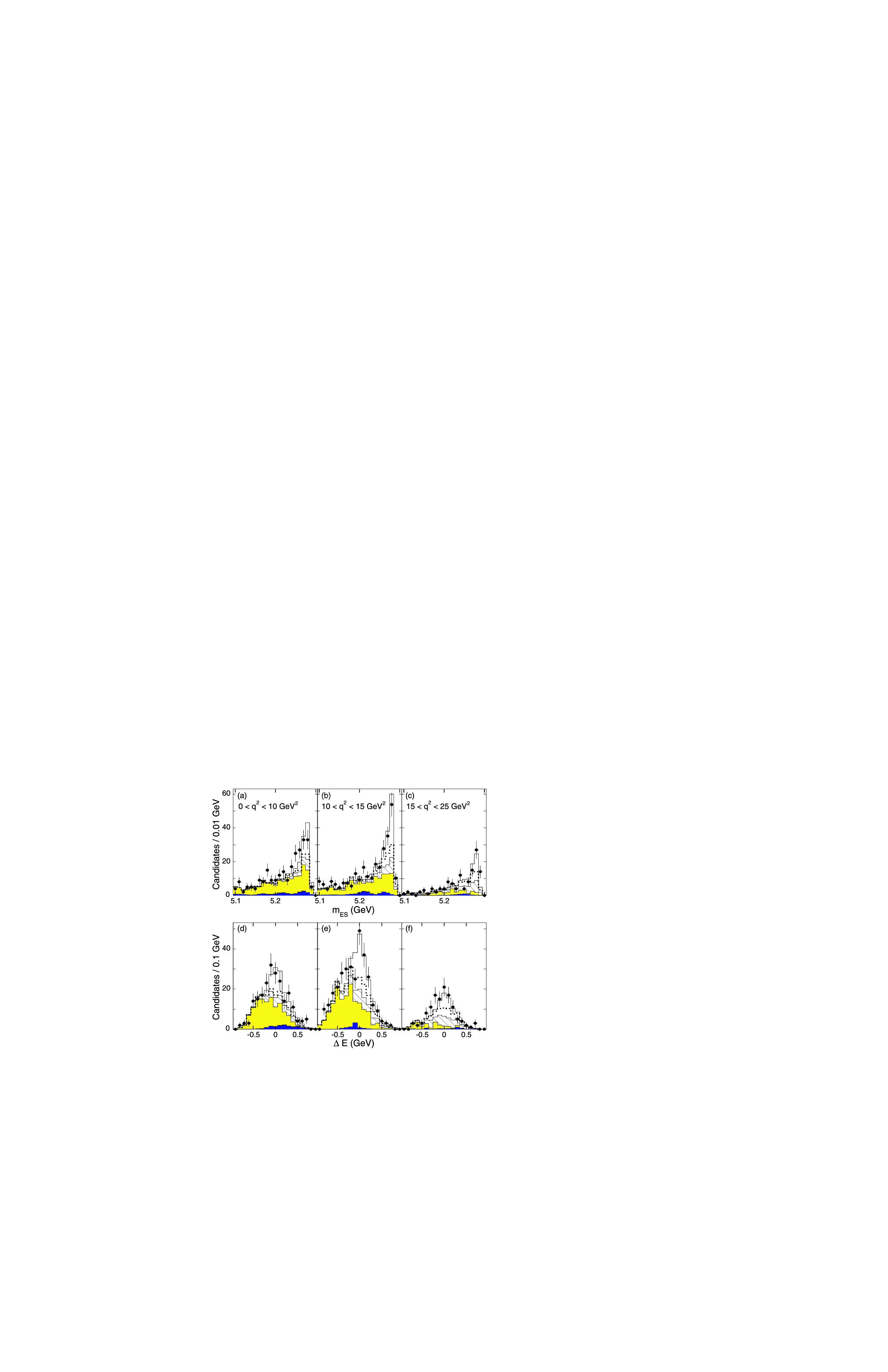, width = 8cm}
\caption{(color online) Projected \mES (a-c) and \DeltaE (d-f) distributions in three intervals 
of $q^2$ for the combined \Btorholnu modes. 
  See the caption for Figure~\ref{fig:MesDeltaE:pi} for details.}
\label{fig:MesDeltaE:rho}
\end{center}
\end{figure}

Figures~\ref{fig:MesDeltaE:pi} and \ref{fig:MesDeltaE:rho} show projections of the fitted 
\DeltaE vs. \mES distributions for each $q^2$ interval for \Btopilnu and \Btorholnu, respectively. 
Integrated over the whole $q^2$ range, we observe 
$396$ \pimlnu, $137$ \pizlnu, $95$ \rhomlnu, and $98$ \rhozlnu decays. 
The resulting partial and total branching fractions are given in Table~\ref{tab:bfq2}.
The fitted normalization of the \bclnu background is
consistent with the measured total branching fraction~\cite{BFbtoc}.
The goodness-of-fit is evaluated using a $\chi^2$-based comparison of the fitted 
$\DeltaE$~vs.~$m_{\rm ES}$ distributions and data, yielding $\chi^2/\rm dof = 1.27$. 
As a check, we have performed the fit for $e^\pm$ and $\mu^\pm$ separately
and obtain consistent results.  

The fit also allows us to study the $q^2$ dependence of the form factors. 
In decays to pseudoscalar mesons there is only one 
form factor, $f_+$ (for low-mass leptons), and we can extract 
the shape of $f_+(q^2)$ directly from the measured $q^2$ spectrum. 
We perform a $\chi^2$ fit with a function proposed by Becirevic 
and Kaidalov (BK)~\cite{bk},
\begin{equation}
  f_+(q^2) = \frac{c_B (1-\alpha)}
                  {(1-q^2/m_{B^*}^2)(1-\alpha q^2/m_{B^*}^2)}\ ,
\end{equation}
where $m_{B^*} = 5.32~\gev$ is the mass of the $B^*$ resonance, $c_B$ is a normalization factor,
and $\alpha$ is a shape parameter. Since we cannot measure the normalization, 
only $\alpha$ is meaningful. Leaving both $c_B$ and $\alpha$ free, 
we fit $\alpha = 0.61 \pm 0.09$, in agreement 
with LQCD results~\cite{lqcd:fnal,lqcd:hpqcd}. 
For decays to vector mesons, there are three form factors. 
The experimental uncertainties for \Btorholnu are still too 
large to measure these. Thus we have to rely on theoretical predictions.

\begin{figure}[htbp]
\begin{center}
  \epsfig{file=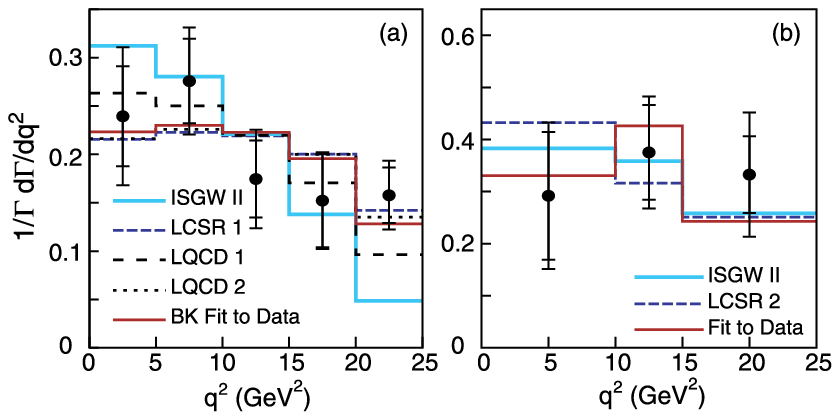,width=\columnwidth}
\caption{(color online) Comparison of the differential decay rates as
  functions of $q^2$ for \Btopilnu~(a) and \Btorholnu~(b) 
  with various form-factor predictions.  The data are background
  subtracted and corrected for efficiency and radiative effects.
  The error bars are statistical (inner) and statistical plus systematic (outer).}
\label{fig:q2Spectra}
\end{center}
\end{figure}

Figure~\ref{fig:q2Spectra} compares the $q^2$ distributions for
\Btopilnu\ and \Btorholnu\ with the various form-factor calculations, 
which we implement by reweighting simulated signal events~\cite{FFReweighting}. 
We use $\chi^2$ probabilities to quantify the agreement:
for \Btopilnu we obtain good agreement 
with the BK fit to the data, $P(\chi^2) = 35\%$;
and the predictions of  
LCSR1~\cite{lcsr:pi}, $38\%$;
LQCD1~\cite{lqcd:hpqcd}, $14\%$;
and
LQCD2~\cite{lqcd:fnal}, $35\%$;
but only marginal agreement with the prediction of ISGW~II~\cite{isgw2}, $P(\chi^2)<1\%$. 
For \Btorholnu all calculations~\cite{lcsr:rho,isgw2}
are compatible with the data within the large experimental uncertainties.

\input{tab_systematics.tex}

The systematic errors in the extraction of the branching fractions 
are listed in Table~\ref{tab:systematics}. The contributions from
each $q^2$ interval are conservatively treated as fully correlated and added
linearly to obtain the uncertainty of the total branching fractions.
Part of the $q^2$~variation of the stated errors may be due to statistical
variations in simulated samples.

Uncertainties in the simulation of the reconstruction of charged particles and photons 
are evaluated by varying the reconstruction efficiencies and the
photon-energy resolution and are added in quadrature. 
In addition, most $K^0_L$ escape detection. 
The impact of $K^0_L$ interactions in the calorimeter is estimated by
varying in simulation their detection 
efficiency and energy deposition. To assess the uncertainty of the $K^0$ production rate, we vary
the inclusive branching fractions of $D^+ \rightarrow \Kbar^0 X$, 
$D^0 \rightarrow \Kbar^0 X$, and $D^+_s \rightarrow \Kbar^0 X$ within their 
published errors~\cite{pdg2004}. All these constitute the total 
uncertainty of the neutrino reconstruction, which is dominant. 
For lepton identification we use relative uncertainties of $\pm 2\%$ and $\pm 4\%$ 
for electrons and muons, respectively.

The uncertainty of the \bclnu background is evaluated by varying the 
$B \rightarrow D / D^* / D^{**} \ell \nu$ branching fractions~\cite{pdg2004} 
and the $B \ra D^* \ell \nu$ form factors~\cite{FFDstar}.
For the \bulnu background, we independently vary the branching fractions of 
\Bpomegalnu  and \Bpetalnu within their published errors~\cite{BFomegalnu, pdg2004}.
We assume equal branching fractions for $\eta\ell\nu$ and $\eta'\ell\nu$ and use a 
relative uncertainty of $100\%$ for the latter. 
We also vary the non-resonant contribution within the range allowed by the
uncertainty of the total \BXulnu branching fraction~\cite{bxulnu}. The
impact of quark-hadron duality violation or weak annihilation effects have
not been considered.  We estimate the uncertainty of the small remaining non-\BB\ background 
by comparing simulation with off-resonance data and extract a normalization
error of ${}_{-25}^{+70} \%$ for electrons and $\pm 25$\% for muons.   

The overall uncertainty of the number of produced $B$ mesons is 1.1\%. 
We take into account the uncertainties of the ratio of \B\ lifetimes, 
\BlifetimeRatio~\cite{pdg2004},
the charged-to-neutral \B\ production ratio \fpmfzz~\cite{pdg2004}, and the potential effect 
of isospin breaking due to $\rho^0-\omega$ mixing~\cite{diazcruz}.  
We assign an uncertainty of 20\% to the radiative corrections based on PHOTOS~\cite{photos}. 

The impact of the uncertainties of the \Btopilnu\ form-factor shape on the 
measured branching fractions is negligible, whereas for the
different \Btorholnu\ form-factor calculations we see variations of 
up to $\pm 6\%$ in \BRBzpilnu and $\pm 13\%$ in \BRBzrholnu. 
We take the full spread between calculations 
as the uncertainty of the $q^2$ dependence of the form factors.

\input{tab_vub.tex}

We extract \Vub (see Table~\ref{tab:vub}) from the partial branching
fractions $\Delta{\cal B}$ using the relation 
$\Vub = \sqrt{\Delta{\cal B}/(\tau_{B^0} \Delta\zeta)}$\,,
where $\tau_{B^0} = (1.536 \pm 0.014)$~ps~\cite{pdg2004} is the $B^0$ lifetime
and $\Delta\zeta$ denotes the predicted form-factor
normalization in each $q^2$ interval.
For $q^2 < 15 \gev^2$ we derive \Vub using LCSR calculations;
for $q^2 > 15 \gev^2$ we use unquenched LQCD.
To extract \Vub from this measurement over the whole $q^2$ range, we extrapolate  
the LQCD results to low $q^2$ using the fits of the BK parametrization in 
Ref.~\cite{lqcd:hpqcd,lqcd:fnal} and the LCSR results to high $q^2$
using the parametrization given in Ref.~\cite{lcsr:pi}. 
We adopt the uncertainties of the form-factor normalization estimated 
in Refs.~\cite{lcsr:pi,lcsr:rho,lqcd:hpqcd,lqcd:fnal}.

In conclusion, we have measured
the exclusive branching fractions \BRBzpilnu and \BRBzrholnu as a function
of $q^2$, and have extracted \Vub using recent form-factor calculations.
We measure the total branching fractions,
\begin{eqnarray*}
  \BRBzpilnu  &=& (1.38 \pm 0.10 \pm 0.16 \pm 0.08)\times 10^{-4} ,\\
  \BRBzrholnu &=& (2.14 \pm 0.21 \pm 0.48 \pm 0.28)\times 10^{-4} ,
\end{eqnarray*}
where the errors are statistical (data and simulation), experimental systematic, 
and uncertainties of the form-factor shapes.
As a consistency check, we have also measured the branching fractions
for the charged and neutral $\pi \ell \nu$ samples separately, 
$\BRBzpilnu  = (1.41 \pm 0.17 \pm 0.17 \pm 0.08) \times 10^{-4}$,
$\BRBppizlnu = (0.70 \pm 0.10 \pm 0.08 \pm 0.04) \times 10^{-4}$.
The ratio
$\Gamma(\Bzpilnu)/\Gamma(\Bppizlnu) = 2.21\pm 0.41$ 
is consistent with the assumed isospin relation within the quoted
statistical uncertainty.

For \Btopilnu, the $q^2$ distribution agrees well with calculations based on 
LCSR~\cite{lcsr:pi} and unquenched LQCD~\cite{lqcd:fnal,lqcd:hpqcd}, 
but the data disfavor ISGW~II~\cite{isgw2}.  
Instead of averaging results based on different
calculations, we choose the measured form-factor shape and normalization of LQCD2 and quote
\begin{eqnarray*}
|V_{ub}| = (3.82 \pm 0.14 \pm 0.22 \pm 0.11 ^{+0.88}_{-0.52}) \times 10^{-3} ,
\end{eqnarray*}
where the additional fourth error reflects the uncertainty of the form-factor
normalization.  The results are consistent with previous
measurements~\cite{cleo_vub,babar_vub}, but have higher
statistical accuracy, are less dependent on theoretical form-factor predictions, 
and benefit from recent advances in theoretical
calculations~\cite{lqcd:fnal,lqcd:hpqcd,lcsr:pi,lcsr:rho}.

We would like to thank P.~Ball, R.~Zwicky, M.~Okamoto, and J.~Shigemitsu  
for their help with theoretical form-factor predictions.  
\input pubboard/acknow_PRL.tex

\bibliography{paper_hepex}

\end{document}

%% file: pubboard/authors_pub05025_weinstein.tex
%
\author{B.~Aubert}
\author{R.~Barate}
\author{D.~Boutigny}
\author{F.~Couderc}
\author{Y.~Karyotakis}
\author{J.~P.~Lees}
\author{V.~Poireau}
\author{V.~Tisserand}
\author{A.~Zghiche}
\affiliation{Laboratoire de Physique des Particules, F-74941 Annecy-le-Vieux, France }
\author{E.~Grauges}
\affiliation{IFAE, Universitat Autonoma de Barcelona, E-08193 Bellaterra, Barcelona, Spain }
% ...
\collaboration{The \babar\ Collaboration}
\noaffiliation

%% file: abstract.tex
%
%

We present an analysis of exclusive charmless semileptonic $B$-meson decays
based on 83 million $B\overline B$ pairs recorded with the 
\mbox{\slshape B\kern-0.1em{\smaller A}\kern-0.1em B\kern-0.1em{\smaller A\kern-0.2em R}} detector at the $\Upsilon(4S)$ resonance.
Using isospin symmetry, we measure branching fractions 
${\cal B}(B^0 \rightarrow \pi^-\ell^+\nu) =
(1.38 \pm 0.10 \pm 0.16 \pm 0.08)\times 10^{-4}$ and ${\cal B}(B^0
\rightarrow \rho^-\ell^+\nu) = (2.14 \pm 0.21 \pm 0.48 \pm 0.28)\times
10^{-4}$, where the errors are statistical, experimental systematic, and
due to form-factor shape uncertainties.
We compare the measured distribution in $q^2$, the momentum-transfer
squared, with theoretical predictions for the form factors from lattice QCD
and light-cone sum rules, and extract the Cabibbo-Kobayashi-Maskawa (CKM) 
matrix element 
$|V_{ub}| = (3.82 \pm 0.14 \pm 0.22 \pm 0.11 ^{+0.88}_{-0.52}) \times 10^{-3}$ 
from $B \rightarrow \pi\ell\nu$, where the fourth
error reflects the uncertainty of the form-factor normalization.

%% file: tab_bfq2.tex
\begin{table}[htbp]
\begin{center}
\caption{Partial and total branching fractions \BRBzpilnu ($\Delta \cal{B}_\pi$) and \BRBzrholnu ($\Delta \cal{B}_\rho$) 
obtained from the simultaneous fit of the four signal modes, and signal efficiencies, $\epsilon_\pi$ and $\epsilon_\rho$, averaged over charged and neutral \B decays.
The errors are statistical.}  
\label{tab:bfq2}
\renewcommand{\arraystretch}{1.2}
\renewcommand{\multirowsetup}{\centering}
\begin{tabular}{r@{\,--\,}lcc@{\qquad}ccc} \hline\hline

  \multicolumn{2}{c}{$q^2$ Range} & $\Delta\cal{B}_\pi$ & $\epsilon_\pi$
   & $q^2$ Range & $\Delta\cal{B}_\rho$ & $\epsilon_\rho$ \\

  \multicolumn{2}{c}{($\gev^2$)} & ($10^{-4}$) & ($\%$)
  &($\gev^2$) & ($10^{-4}$) & ($\%$) \\ \hline

  \ 0&5      & $0.30\pm0.05$ & \ $2.1$ & \multirow{2}{.4in}{\ 0\,--\,10}
                             & \multirow{2}{.7in}{$0.73\pm0.17$} & \multirow{2}{.2in}{$0.70$}\\

  \ 5&10     & $0.32\pm0.05$ & \ $2.9$ & & \\

  \ 10&15    & $0.23\pm0.05$ & \ $3.8$ &10\,--\,15 & $0.82\pm0.10$ & $0.97$\\

  \ 15&20    & $0.27\pm0.05$ & \ $3.5$ & \multirow{2}{.4in}{15\,--\,25} 
                             & \multirow{2}{.7in}{$0.59\pm0.07$} & \multirow{2}{.2in}{$0.44$}\\

  \ 20&25    & $0.26\pm0.03$ & \ $3.3$ & &\\

  \  0& 25   & $1.38\pm0.10$ & \ $3.1$ &\  0\,--\,25 & $2.14\pm0.21$ & $0.72$ \\
\hline\hline
\end{tabular}
\end{center}
\end{table}

%% file: tab_systematics.tex
\begin{table*}[htbp]
\begin{center}
\caption{Relative systematic uncertainties of the partial and total branching
  fractions \BRBzpilnu ($\Delta \cal{B}_\pi$) and \BRBzrholnu ($\Delta \cal{B}_\rho$) in the
  various $q^2$ bins.  The total uncertainty in each column is the sum in
  quadrature of the listed contributions.}
\label{tab:systematics}
\renewcommand{\arraystretch}{1.2}
\begin{tabular}{c|cccccc|cccc} \hline\hline
 & \multicolumn{6}{c|}{$\delta \Delta {\cal{B}_{\pi}} / \Delta \cal{B}_{\pi}$ ($\%$)} & \multicolumn{4}{c}{$\delta \Delta {\cal{B}_{\rho}} / \Delta \cal{B}_{\rho}$ ($\%$)} \\
$q^2$ Range ($\gev^2$)               &0--5&5--10 &10--15&15--20&20--25& 0--25
                                     &0--10&10--15&15--25& 0--25 \\ \hline
Track and Photon Reconstruction      & 7.4 & 5.7 & 9.2 & 3.5 & 8.7 & 6.8 &16.7 &10.8 &15.9 &14.2 \\ 
$K^0_L$ Production and Interactions  & 8.8 & 5.0 & 7.3 & 2.5 & 4.8 & 5.7 &12.7 & 4.7 &10.9 &9.1  \\
Lepton Identification                & 2.3 & 2.3 & 2.3 & 2.3 & 2.3 & 2.3 & 2.3 & 2.3 & 2.3 & 2.3 \\ \hline
$B\rightarrow X_c\ell\nu$ Background & 5.0 & 3.9 & 3.9 & 4.3 & 3.6 & 4.2 & 7.2 & 1.8 & 3.8& 4.2 \\ 
$B\rightarrow X_u\ell\nu$ Background & 0.5 & 1.5 & 0.6 & 2.2 & 5.7 & 2.1 & 10.9 & 9.1 & 19.2& 12.5\\ 
Non-\BB\ Background	             &13.5 & 2.4 & 1.0 & 2.2 & 7.8 & 5.6 &11.2 & 0.9 & 1.6& 4.6 \\ \hline
$N_{\BB}$                            & 1.1 & 1.1 & 1.1 & 1.1 & 1.1 & 1.1 & 1.1 & 1.1 & 1.1& 1.1 \\ 
\B Lifetimes                         & 1.1 & 1.4 & 0.2 & 0.5 & 2.1 & 1.1 & 0.7 & 0.6 & 0.9& 0.7 \\ 
$f_{+-}/f_{00}$                      & 0.7 & 0.8 & 0.3 & 0.4 & 1.0 & 0.7 & 1.6 & 0.6 & 0.4& 0.9 \\ 
Isospin Breaking                     & 0.1 & 1.1 & 1.8 & 0.1 & 0.1 & 0.6 & 6.4 & 4.5 & 0.9& 4.2 \\
Radiative Corrections                & 0.8 & 0.3 & 0.2 & 0.3 & 1.5 & 0.6 & 0.5 & 0.1 & 0.9& 0.5 \\ \hline
Total Error                          &18.7 &  9.5 &12.8 & 7.3 &14.8 &11.8 &28.1 &15.9 &27.7&22.5 \\ \hline\hline
\end{tabular}
\end{center}
\end{table*}

%% file: tab_vub.tex
\begin{table}[thbp] 
  \centering
  \caption{$|V_{ub}|$ derived for \Btopilnu and \Btorholnu signal for 
  various $q^2$ regions and form-factor (FF) calculations:
  LCSR1~\cite{lcsr:pi}, LQCD1~\cite{lqcd:hpqcd}, LQCD2~\cite{lqcd:fnal}, LCSR2~\cite{lcsr:rho}, ISGW~II~\cite{isgw2}. 
  For the cross feed from the other mode, we have used the 
  BK fit to data for \pilnu and LCSR2 for \rholnu.
  Quoted errors are statistical,
  experimental systematic, uncertainties of form-factor shape and
  form-factor normalization~$\Delta\zeta$
  (no form-factor normalization uncertainties are available for \rholnu).} 
  \label{tab:vub}
  \renewcommand{\arraystretch}{1.2}
  \begin{tabular}{l@{\ }r@{\,--\,}lc@{\quad}l} \hline\hline
  & \multicolumn{2}{c}{$q^2$ Range} & $\Delta\zeta$ &
    \multicolumn{1}{c}{$|V_{ub}|$} \\
  & \multicolumn{2}{c}{($\gev^2$)} & (ps$^{-1}$) &
    \multicolumn{1}{c}{(10$^{-3}$)} \\
\hline
  $\pi$ FF \\
  LCSR1   &\ 0&15  &  $5.1{\pm}1.3$ & $3.27 \pm 0.16 \pm 0.19 \pm 0.10 ^{+0.53}_{-0.36}$ \\
  LQCD1   &\ 15&25 &  $1.5{\pm}0.4$ & $4.92 \pm 0.25 \pm 0.29 \pm 0.15 ^{+0.76}_{-0.52}$ \\
  LQCD2   &\ 15&25 &  $2.0{\pm}0.5$ & $4.16 \pm 0.22 \pm 0.24 \pm 0.12 ^{+0.72}_{-0.47}$ \\
  LCSR1   &\ 0&25  &  $7.7{\pm}2.3$ & $3.40 \pm 0.13 \pm 0.20 \pm 0.10 ^{+0.67}_{-0.42}$ \\
  LQCD1   &\ 0&25  &  $5.7{\pm}1.7$ & $4.00 \pm 0.14 \pm 0.23 \pm 0.12 ^{+0.78}_{-0.49}$ \\
  LQCD2   &\ 0&25  &  $6.1{\pm}2.1$ & $3.82 \pm 0.14 \pm 0.22 \pm 0.11 ^{+0.88}_{-0.52}$ \\
\hline
  $\rho$ FF \\
  LCSR2   &\ 0&15  & $12.7$  & $2.82 \pm 0.18 \pm 0.30 \pm 0.18 ^{}_{}$ \\
  ISGW~II &\ 0&25  & $14.2$  & $2.91 \pm 0.12 \pm 0.33 \pm 0.19 ^{}_{}$ \\
  LCSR2   &\ 0&25  & $17.2$  & $2.85 \pm 0.14 \pm 0.32 \pm 0.19 ^{}_{}$\\
\hline\hline
  \end{tabular}
\end{table}

%% file: pubboard/acknow_PRL.tex
We are grateful for the excellent luminosity and machine conditions
provided by our \pep2\ colleagues, 
and for the substantial dedicated effort from
the computing organizations that support \babar.
The collaborating institutions wish to thank 
SLAC for its support and kind hospitality. 
This work is supported by
DOE
and NSF (USA),
NSERC (Canada),
IHEP (China),
CEA and
CNRS-IN2P3
(France),
BMBF and DFG
(Germany),
INFN (Italy),
FOM (The Netherlands),
NFR (Norway),
MIST (Russia), and
PPARC (United Kingdom). 
Individuals have received support from CONACyT (Mexico), A.~P.~Sloan Foundation, 
Research Corporation,
and Alexander von Humboldt Foundation.